\documentstyle[12pt,psfig]{article}
	\newlength{\paperbaselineskip}
\newfont{\fourteencp}{cmcsc10 scaled\magstep2}
\newfont{\titlefont}{cmbx10 scaled\magstep2}
\newfont{\authorfont}{cmcsc10 scaled\magstep1}
\newfont{\fourteenmib}{cmmib10 scaled\magstep2}
	\skewchar\fourteenmib='177
\newfont{\elevenmib}{cmmib10 scaled\magstephalf}
	\skewchar\elevenmib='177
\newfont{\ninemib}{cmmib9} \skewchar\ninemib='177
\makeatletter
\newif\ifpr@pstyle \pr@pstylefalse
\newif\ifnons@qeq  \nons@qeqfalse
\newcommand\nonsequentialeqnum{
        \nons@qeqtrue
	\@addtoreset{equation}{section}
	\def\theequation{\arabic{section}.\arabic{equation}}}
\newif\ifp@bblock  \p@bblocktrue
\newcommand\nopubblock{\p@bblockfalse}
\newcommand\topspace{\hrule height 0pt depth 0pt \vskip}
\newcommand\p@bblock{\begingroup \tabskip=\hsize minus \hsize
	\baselineskip=1.5\ht\strutbox \topspace-2\baselineskip
	\halign to\hsize{\strut ##\hfil\tabskip=0pt\crcr
	\the\Pubnum\crcr\the\date\crcr}\endgroup}
\newcommand\YUKAWAmark{\hbox{
        \ifpr@pstyle\ninemib\else\elevenmib\fi
        Yukawa\hskip1mm Institute\hskip1mm Kyoto \hfill}}
\newtoks\date
\newtoks\Pubnum

\newcommand{\frontpageskip}{\vspace{12pt plus .5fil minus 2pt}}
\def\@authoraddress{} \def\@title{}
\def\title#1{\gdef\@title{\frontpageskip
	\begin{center}{\titlefont #1}\end{center}\par}}
\def\@author#1{\frontpageskip\par\begin{center}{\authorfont #1}
	\end{center}
	\nobreak}
\def\author#1{\expandafter\def\expandafter\@authoraddress\expandafter
    {\@authoraddress{\@author{#1}}}}
\def\andauthor#1{\expandafter\def\expandafter\@authoraddress\expandafter
    {\@authoraddress{\frontpageskip\centerline{and}\@author{#1}}}}
\def\authors#1{\expandafter\def\expandafter\@authoraddress\expandafter
    {\@authoraddress{\frontpageskip\noindent #1}}}
\def\@address#1{\par\begin{center}{\sl #1}\end{center}\par}
\def\address#1{\expandafter\def\expandafter\@authoraddress\expandafter
    {\@authoraddress{\@address{#1}}}}
\def\andaddress#1{\expandafter\def\expandafter%
    \@authoraddress\expandafter
    {\@authoraddress{\par\centerline{\sl and}\@address{#1}}}}
\renewcommand{\thanks}[1]{\footnote{#1}}
\def\maketitle{\par
  \begingroup
       \def\thefootnote{\fnsymbol{footnote}}
	\thispagestyle{empty}
        \baselineskip=\paperbaselineskip
	\@maketitle
	\endgroup
	\setcounter{footnote}{0}
	\let\maketitle\relax \let\@maketitle\relax
	\let\@thanks\relax \let\@title\relax
	\let\@title\relax \let\@authoraddress\relax
	\let\thanks\relax}
\def\@maketitle{%
        \ifpr@pstyle\vspace{-1.0cm}\else\vspace{-1.7cm}\fi
	\YUKAWAmark\vskip0.6cm
	\ifp@bblock\p@bblock \else\hrule height 0pt \relax \fi
	\@title
	\@authoraddress
	}
\renewcommand{\abstract}{\par\frontpageskip\centerline{
             \ifpr@pstyle\twelvecp\else\fourteencp\fi Abstract}
	\vspace{8pt plus 3pt minus 3pt}}
\makeatother
\newcommand{\Gammarot}{$\Gamma_{rot}$}
\newcommand{\Gammamu}{$\Gamma_\mu$}
\newcommand{\Uonset}{$U_{onset}$}
\newcommand{\Nband}{$N_{band}$}
\newcommand{\Yb}{${}^{168}$Yb}
\newcommand{\Te}{${}^{114}$Te}
\newcommand{\U}{${}^{234}$U}
\newcommand{\Eu}{${}^{143}$Eu}
\newcommand{\Dy}{${}^{152}$Dy}
\newcommand{\Hg}{${}^{192}$Hg}
\newcommand{\eps}{\epsilon}
\newcommand{\epsh}{\epsilon_4}
\def\m@thcombine#1#2{%
  \setbox0=\hbox{$#1$}
  \setbox1=\hbox{$#2$}
  \ifdim\wd0>\wd1
    \setbox0=\hbox to\wd1{\hss\box0\hss}
  \else
    \setbox1=\hbox to\wd0{\hss\box1\hss}
  \fi
  \mathop{\vcenter{
    \offinterlineskip\box0\box1}}}
\def\lesim{\m@thcombine<\sim}
\def\gesim{\m@thcombine>\sim}

\Pubnum={YITP-98-68}
\date={October 1998}

\title{\Large\bf Microscopic Structure of Rotational Damping 
\thanks{A talk presented by M.M. at the {\it Topical 
Conference on Giant Resonances}, Varenna, May 11-16, 1998}}

\author{M. Matsuo${}^a$,
K. Yoshida${}^b$,
T. D\o ssing${}^c$,
E. Vigezzi${}^d$,
R.A. Broglia${}^{c,d}$}
\address{\small\sl ${}^a$Yukawa Institute for Theoretical Physics, 
	Kyoto University, Kyoto 606-8502, Japan}
\address{\small\sl ${}^b$Research Center for Nuclear Physics, 
	Osaka University, Osaka 567-0047, Japan}
\address{\small\sl ${}^c$Niels Bohr Institute, University of Copenhagen,
	DK2100 Copenhagen \O, Denmark}
\address{\small\sl ${}^d$INFN Sez. Milano, and Department of Physics, 
	University of Milano, Milan 20133, Italy}

\begin{document}
\maketitle

\begin{abstract}
The damping of collective rotational motion is studied microscopically,
making use of  shell model calculations
based on the cranked Nilsson deformed mean-field and on residual two-body
interactions, and focusing on the shape of the gamma-gamma correlation 
spectra and on its systematic behavior. It is shown that the spectral
shape is directly related to 
the damping width   of collective rotation, \Gammarot, 
and to the spreading width
  of many-particle many-hole configurations, \Gammamu.
The rotational damping width is affected  by
the shell structure, and is very sensitive 
to the position of the Fermi surface,
besides mass number,  spin and deformation.
This produces a  rich variety of features in the rotational damping phenomena.
\vspace{20mm}
\end{abstract}

\section{Introduction}

The collective rotation of deformed nuclei becomes a damped motion as
the nuclei are thermally excited \cite{Leander,Lauri86,FAM,RPM}.
It is known that the levels near the yrast line form
rotational band structures based on simple 
configurations which are well described by cranked mean-field models.
As the thermal excitation energy ( i.e., the excitation
energy measured from the yrast line) increases, the level density
also increases  steeply, so that the  spacings between neighboring levels 
become smaller than the typical size
($\sim 10$ keV) of the two-body matrix elements of the residual
interaction.
In such a situation, the pure configurations of the cranked mean-field 
get mixed and form compound states.
Because of this configuration mixing, the rotational E2 decay
from an off-yrast compound state (at spin $I$) may be fragmented over 
many final
states (at $I-2$). The width of the associated strength function 
corresponds to the rotational damping width \Gammarot, that is, to
the damping width of collective rotation.

A microscopic description 
of rotational damping is provided by
a shell model diagonalization using 
the cranked Nilsson-Strutinsky mean-field
and a two-body residual interaction such as the surface delta 
interaction (SDI) or the delta force \cite{Aberg,Matsuo93,Matsuo97}. 
It is important
to include all many-particle many-hole (np-nh) configurations present in
the energy region of interest in order to enable the 
shell model description of the off-yrast energy levels and of the associated
rotational E2 transitions. 
This is possible since the intrinsic
excitation energy which is relevant for the observed gamma-ray
spectra is not very high (up to a few MeV), corresponding
to about $10^2-10^3$ levels for each spin and parity in heavy nuclei, a number
which can be handled by standard diagonalization techniques.
Thus one can perform detailed studies of the microscopic
structure of the damping of collective rotation.
In this paper, we report our recent theoretical investigations, 
particularly concerning the spectral shape of damped rotational E2
transitions and its systematic behavior.

\section{Rotational damping and doorway states of damped E2 transitions}

\begin{figure}[h]
\centerline{\psfig{figure=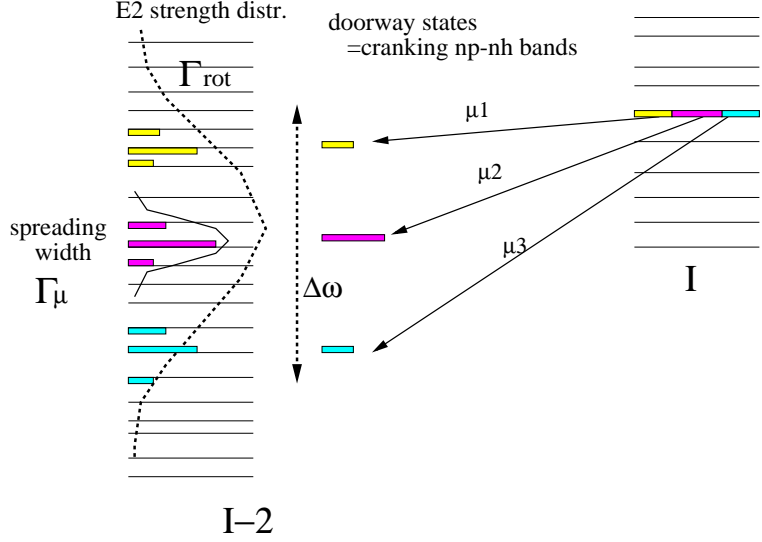,width=100mm,angle=0}}
\caption{Schematic illustration of 
damped rotational E2 transitions in rotating nuclei. The horizontal
bars represent the energy levels at spin $I$ and $I-2$. The thick
bars  represent the strength of cranked shell model np-nh states.
See also text.}
\end{figure}

Using the cranked shell model \cite{Lauri86,Aberg,Matsuo97}, 
the energy eigenstates (energy levels) at spin $I$ are
described as linear superpositions
of basis cranked np-nh configurations $\left|\mu(I)\right>$

\begin{equation}
\left|\alpha(I)\right> = \sum_\mu X_\mu^\alpha(I)  \left|\mu(I)\right>
\end{equation}
where the amplitude $X_\mu^\alpha(I)$
 is determined by diagonalization of a shell model
Hamiltonian and the basis space includes all np-nh configurations 
$\left|\mu(I)\right>$
of the cranked Nilsson single-particle orbits.
It is assumed that if 
the residual interaction were not present,
the $\mu$ configurations would form rotational
band structures, so that rotational E2 transitions would connect
only states with the same configuration, with transition energy
 $E_{\gamma \mu} = E_\mu(I)-E_\mu(I-2) =  2 \omega_\mu(I)$.
The compound state $\left|\alpha(I)\right>$, however, contains
many $\mu$ states which in general have different transition energies  
$E_{\gamma \mu}$
(or different rotational frequency $\omega_\mu(I)$)
because of the difference in the angular momentum alignments
of nucleons occupying different orbits. The $\mu$ states
at spin $I-2$ can then be regarded as doorway states for the damped
rotational E2 decay from $\alpha(I)$. Since the doorway states are
not energy eigenstates, they spread over compound states at $I-2$
(due to the residual two-body interaction) with spreading width
\Gammamu.  If the spreading width \Gammamu\  is small,
the total E2 strength distribution reflects the distribution of
doorway states, and its width,
the rotational damping width \Gammarot, 
is proportional to 
the statistical dispersion $\Delta\omega$ of the rotational
frequency $\omega_\mu$ of cranked shell model np-nh states. 
In turn, the doorway structures are smeared 
by the the spreading width \Gammamu\ (See Fig.1), determining the fine
structure of the strength distribution.
If \Gammamu\ is larger than $\Delta\omega$ (the strong
coupling limit), the doorway picture is lost, and one 
is in the region of the ``motional narrowing'' of the damping width.
In this limit,  the rotational damping width is estimated to
be $\Gamma_{rot} \propto \Delta\omega^2/\Gamma_\mu$ 
\cite{Lauri86}.

\begin{figure}[t]
\centerline{\psfig{figure=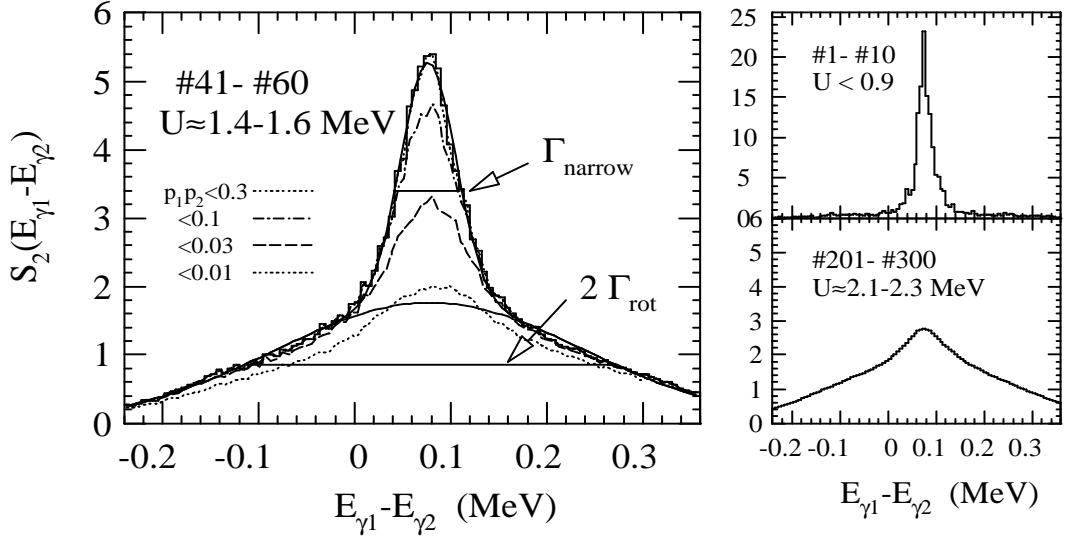,width=14cm,angle=-90}}
\caption{Calculated gamma-gamma correlation spectra for
two consecutive E2 gamma rays
$ I+2 \rightarrow I$ ($E_{\gamma 1}$) and 
$ I \rightarrow I-2$ ($E_{\gamma 2}$), projected on the axis
$E_{\gamma 1}- E_{\gamma 2}$
for the levels of \Yb\ in the
energy bins including the fist to 10th lowest levels (for each
$I^\pi$), the 21-st to 40th, and the 250th to 300th covering
different intrinsic excitation energy $U$ in spin interval $I=31-50$.
}
\end{figure}

For analyses of the rotational damping, it is useful to look at 
two coincident gamma-rays emitted in the E2 decay cascade
\cite{FAM,RPM}.
If there is no rotational damping,
the rotational E2 transition takes place along the rotational bands,
and the energy difference between two consecutive gamma's 
for $I+2 \rightarrow I$ and
for $I \rightarrow I-2$ forms a sharp 
peak at $E_{\gamma 1}-E_{\gamma 2}=4/\cal{J}$ because
of the rotational correlation $E_\gamma \sim 2I/\cal{J}+{\rm
const.}$ ($\cal{J}$ being the rotational 
moment of inertia). 
For E2 transitions associated with the damped rotation, 
on the other hand,
the gamma-gamma correlation spectrum shows a broadening
around the same position, and the width of the correlation 
spectrum can then be related to the rotational damping width 
\Gammarot\ (\Gammarot=FHWM/2 by assuming a Gaussian distribution for the
strength associated to a
single transition step).
The calculated spectra clearly show the change from the region of unperturbed 
rotational bands close to yrast, to the region of damped collective rotation,
as a function of the intrinsic excitation energy $U$ (see Fig.2 and 
Ref.\cite{Matsuo97}).
It should be noted, however, that the spectrum for the region of
$U = 1-2$ MeV, which is most relevant to the experimental quasicontinuum
gamma-ray spectra, shows neither the sharp peak associated with 
the band structure nor the broadened damped distribution. It has  instead
a more complex profile, which can be decomposed in the sum of 
two contributions, having wide and narrow distributions,
as shown in Fig.~2 \cite{Matsuo97,scar}.

\section{The narrow component and spreading width of np-nh states}

To understand the two-component profile, it is useful to 
reconsider the illustration of the rotational damping
shown in Fig.1, where the
E2 strength distribution is characterized not only by 
the damping width \Gammarot\, but also by the spreading width \Gammamu\ of
the np-nh configurations. We can consider the same doorway states
for feeding transitions from states at $I+2$ to the state $\alpha$
at $I$ as those decaying from $\alpha$ at $I$ to $I-2$. Taking the
correlation between feeding and decaying transitions, 
the coincident transitions sharing 
the same doorway states ($\mu$ states) keep
the rotational correlation up to the energy scale of \Gammamu,
and their contribution form the narrow component of width $\sim$ \Gammamu\
around $E_{\gamma 1}-E_{\gamma 2}=4/\cal{J}$.

Indeed the calculations show a direct relation \cite{Spreading} between 
the spreading width \Gammamu\ and the 
width of the narrow component $\Gamma_{narrow}$.
The latter can be extracted from
the calculated gamma-gamma correlation spectra by a fit based on two gaussians
(See Fig.2).
On the other hand, the spreading width \Gammamu\ of the
np-nh shell model configurations $\mu$ is defined in terms of
the strength function 
$S_\mu(E)=\sum_\alpha \left|X_\mu^\alpha\right|^2 \delta(E-E_\alpha)$ 
of the $\mu$ states which are embedded
in the eigenstates $\alpha$'s\cite{BM}. 
The spreading width \Gammamu\ of $\mu$ states 
can be extracted 
by taking the autocorrelation of $S_\mu(E)$,
then averaging over many $\mu$ states, and identifying a half of FWHM 
of the averaged autocorrelation function 
$C(e)=\left<\int dE S_\mu(E)S_\mu(E+e)\right>$ as the spreading
width \Gammamu. 
Figure 3 compares the extracted values of $\Gamma_{narrow}$
and \Gammamu\ for different kinds and force strengths of the residual 
two-body interaction. There exists a clear and universal linear relation
between the two widths.

\begin{figure}[t]
\begin{minipage}[t]{75mm}
\psfig{figure=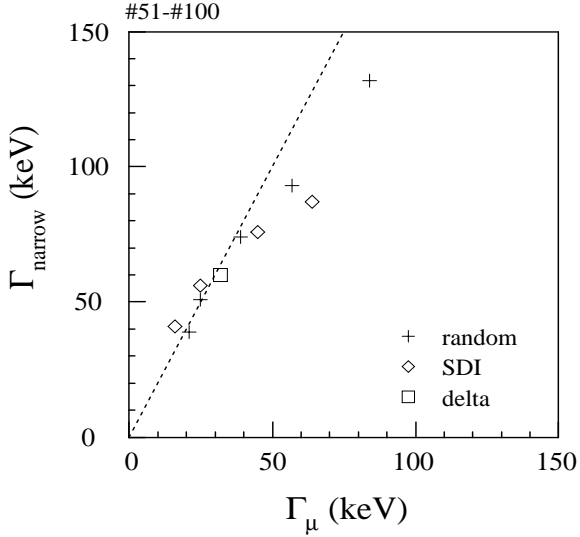,width=75mm,angle=-90}
\end{minipage}
\hspace{\fill}
\begin{minipage}[t]{70mm}
\vspace{-70mm}
\caption{The width of the narrow component $\Gamma_{narrow}$
of the gamma-gamma correlation spectra versus
the spreading width \Gammamu\ of np-nh cranked shell model states,
calculated for the energy bin including 51st to 100th levels
(for each $I^\pi$) in the spin interval $I=20-50$ in \Yb\ using
three different residual two-body interactions (a random two-body
int., the surface delta force, and the volume delta force) with
various force strengths. 
}
\end{minipage}
\end{figure}

\section{Systematics and varieties of rotational damping }

The microscopic calculations predict a large variation 
of rotational damping as a function of deformation,
mass number and nuclear species. 
To demonstrate it, we show in Fig.4
the calculated gamma-gamma correlation spectra for
several normal and
superdeformed nuclei in different mass regions.
Here \Te,\Yb, and \U\ represent typical normal deformed nuclei in
different mass regions.  For \Te, quasicontinuum gamma-ray
spectra from a fusion reaction suggest the presence of
collective rotation in the high spin region $I \gesim 35\hbar$\cite{Tequasi}.
Rotational band structures
were observed recently up to spin  $I \sim 50\hbar$;
the moment of inertia associated with the observed rotational bands 
decreases with spin, indicating that the collective rotation may terminate
at the highest spins \cite{Teband}. In the present calculation for
\Te, we neglect the possibility of band
termination and choose a fixed value for the deformation,
for simplicity.
\U\ is chosen as a typical deformed nucleus in the actinide
region although it may not be  very easy to feed very high spins
because of the competition with fission.
\Eu, \Dy, and \Hg\ are representative of superdeformed nuclei
in $A\sim 150$ and $190$ mass regions. For those nuclei,
quasicontinuum E2 gamma-rays associated with superdeformed states
are observed \cite{Eucont,Schiffer,Lauritsen92,Hgcont}.  
The deformation
parameters used in the calculation are $(\eps,\epsh)=(0.25,0.0),(0.255,0.014),
(0.226,-0.05)$ for   \Te,\Yb, \U\ respectively, and the same values as
in Refs.\cite{Yoshida97,Yoshida98} for superdeformed 
\Eu, \Dy, and \Hg. For \Te, we used a
modified  Nilsson parameter\cite{Zhang}.  As the residual
interaction, the surface delta interaction and the volume delta force 
with standard parameters \cite{Matsuo97} are used for normal and superdeformed
nuclei, respectively. See Refs.\cite{Matsuo97,Yoshida97,Yoshida98}
for further details.

\begin{figure}[t]
\centerline{
\psfig{figure=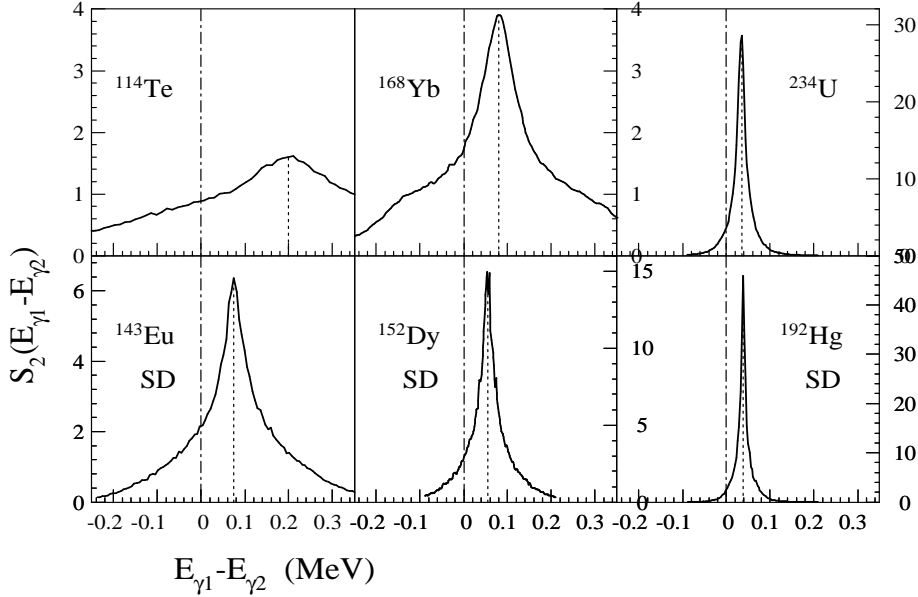,width=12cm,angle=-90}
}
\caption{
The gamma-gamma correlation spectra for
normal deformed ${}^{114}$Te, ${}^{168}$Yb,
 ${}^{234}$U, and superdeformed ${}^{143}$Eu,${}^{152}$Dy,
${}^{192}$Hg.  The spectrum is
calculated for the states from 51st to 100th 
(for each $I^\pi$) in the spin region $I=36-45$ ($I=46-55$ for superdeformed
\Dy\ and \Eu).
}
\end{figure}
\begin{table}[b]
\caption{The calculated rotational damping width \Gammarot\ and
the spreading width \Gammamu\ cranked shell model np-nh  states
for \Te, \Yb, \U, \Eu, \Dy\ and \Hg.
The same energy bin and spin interval as in Fig.4 are considered.
} 
\begin{center}
\begin{tabular*}{15cm}{@{}l@{\extracolsep{\fill}}rrrrrr}
\hline
                 & \Te  & \Yb  & \U 
& \Eu(SD) & \Dy(SD) &  \Hg(SD)  \\
\hline
\ \ \ \Gammarot \ \  (keV) \hspace{1cm}  &  387 & 224 & 33 & 143 & 67& 
23  \\
\ \ \ \Gammamu  \ \  (keV) \hspace{1cm}  &  98  & 45 & 46 &  43 & 59 &
28  \\
\hline
\end{tabular*}
\end{center}
\end{table}

The results shown in Fig.~4 present
significantly different  gamma-gamma
correlation spectra in different
nuclei. Qualitatively, the width of the spectrum 
decreases with increasing mass number and
deformation, in agreement with an analytic estimate by
Lauritzen et al. \cite{Lauri86}. 
The rotational damping width extracted from the gamma-gamma
correlation spectra by the method of the two gaussian fit (See Fig.2)
is summarized in Table 1. 
The damping width in SD \Hg\ is as small as 20 keV, that is, almost
10 times smaller   than in the typical rare-earth normal
deformed nucleus \Yb.

\begin{figure}[t]
\centerline{
\psfig{figure=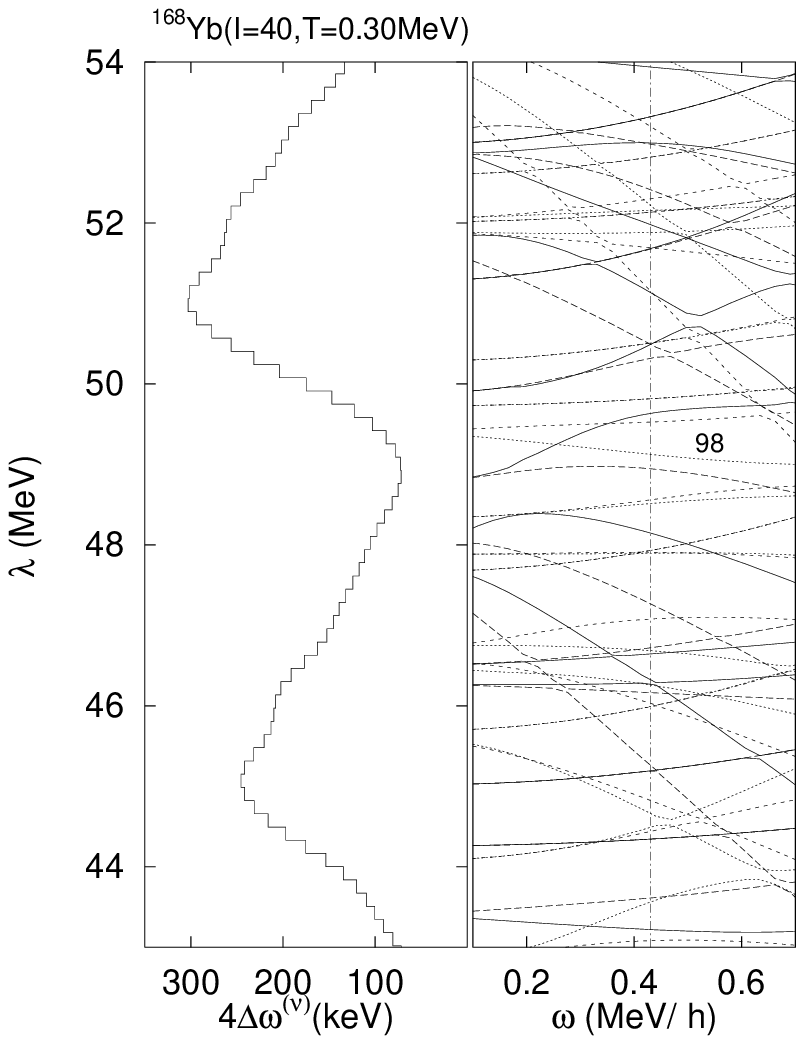,width=7.0cm,angle=0}
\psfig{figure=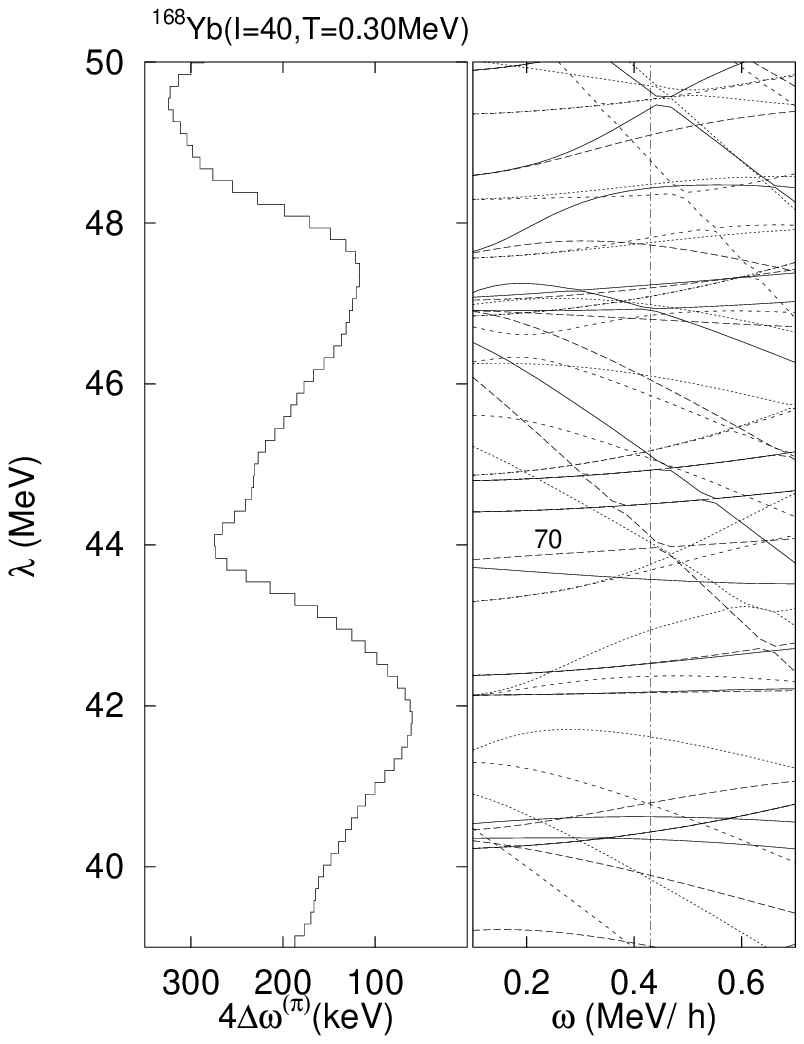,width=7.0cm,angle=0}
}
\caption{The cranked Nilsson single-particle levels as
a function of rotational frequency $\omega$ for \Yb\ for neutrons
(left panel) and protons (right panel), are plotted in the right part of each
panel. In the 
left part of each panel, the neutron and proton contribution to the dispersion
of rotational frequency $\Delta\omega$ is plotted as a function of
the Fermi energy $\lambda$ for $\omega=0.434$ MeV and $T=0.3$ MeV,
corresponding to $I=40, U\sim 2$ MeV}
\end{figure}

The systematic behavior can be understood microscopically.
The rotational damping width \Gammarot\ is dominated 
by the dispersion of rotational frequency $\Delta\omega$
as illustrated in Fig.1. The dispersion $\Delta\omega$
arises from the particle alignments along the rotational
axis. The alignments are different depending on the
cranked single-particle orbits; some are highly alignable
(steeply down-sloping as $\omega$ increases),
while some show small alignments as shown in Fig.~5.
The highly aligned orbits are intruder
orbits with high-$N$ and low-$\Omega$, and therefore
they are located in a specific region in a shell of the cranked single-particle
spectrum. This shell structure 
causes  the dispersion $\Delta\omega$ of rotational
frequency to oscillate strongly as a function of
the position of the Fermi surface,
in addition to the background smooth dependence \cite{Lauri86} 
$\Delta\omega \propto I A^{-5/2}\eps^{-1}$
on spin $I$, quadrupole deformation
$\eps$, and the mass number $A$. Figure 5 also
plots the value of
$\Delta\omega $ as a function of the Fermi energy $\lambda$, 
evaluated microscopically as 
$\Delta\omega=\frac{1}{{\cal J}}\sqrt{\sum_n
i_n^2f_n(1-f_n)}$ \cite{Nishinomiya} 
using the single-particle alignments $i_n$ and
the Fermi-Dirac thermal distribution  
$f_n=(1+\exp{\frac{e_n-\lambda}{T}})^{-1}$ . One clearly sees a shell
oscillation pattern with a large amplitude:
maximum and minimum differ by more than a factor 2.
It is noted  that the difference in the rotational damping width
\Gammarot\ between
SD \Eu\ and SD \Dy\ is considerable in spite of the 
small  difference  in 
deformation and mass number. This arises from the
shell structure of the single-particle alignments, which is 
also responsible for the extremely small value of \Gammarot\ in SD \Hg\
 \cite{Yoshida98}.

\begin{figure}[t]
\centerline{\psfig{figure=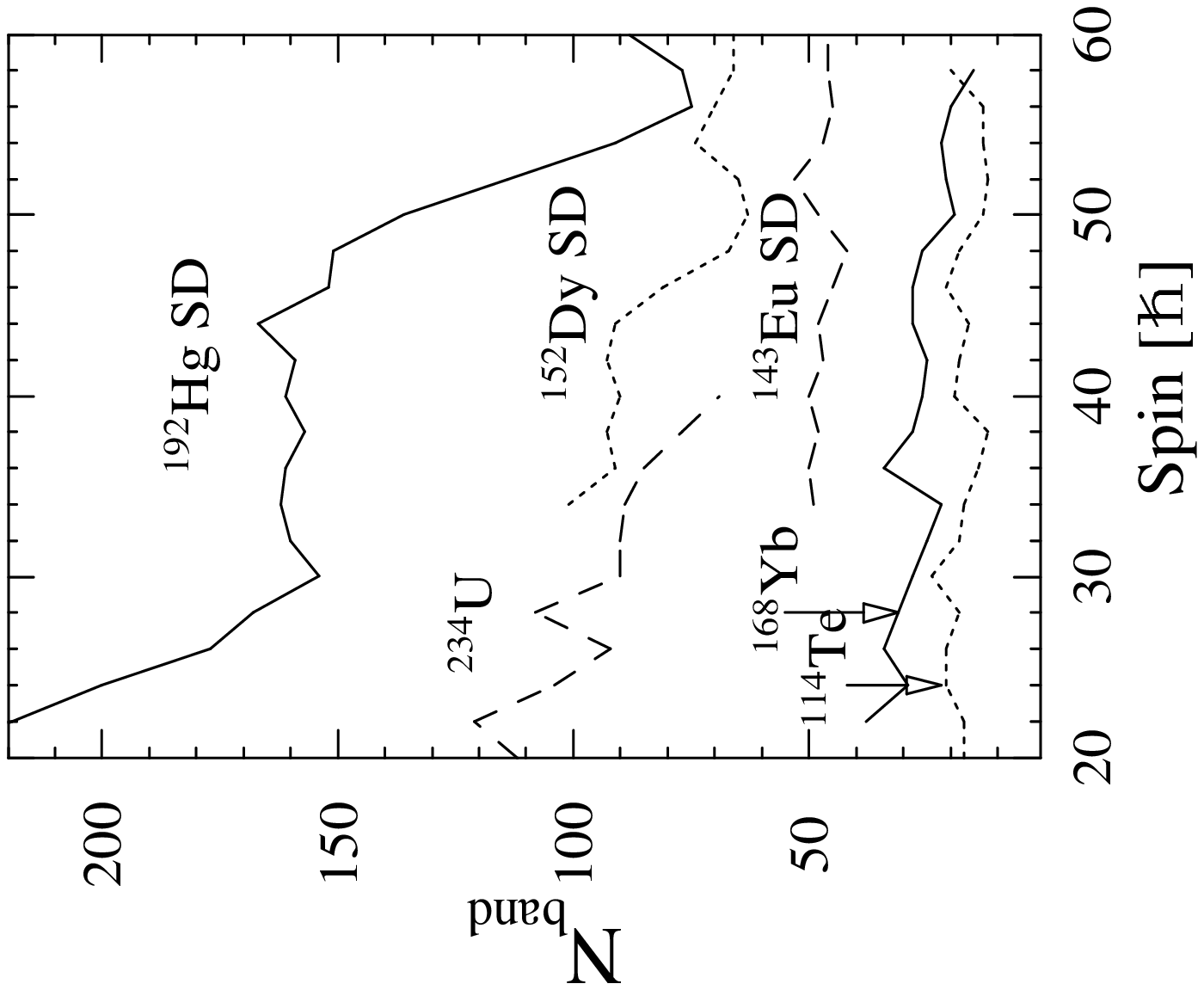,height=7.5cm,angle=-90}
\hspace{5mm}\psfig{figure=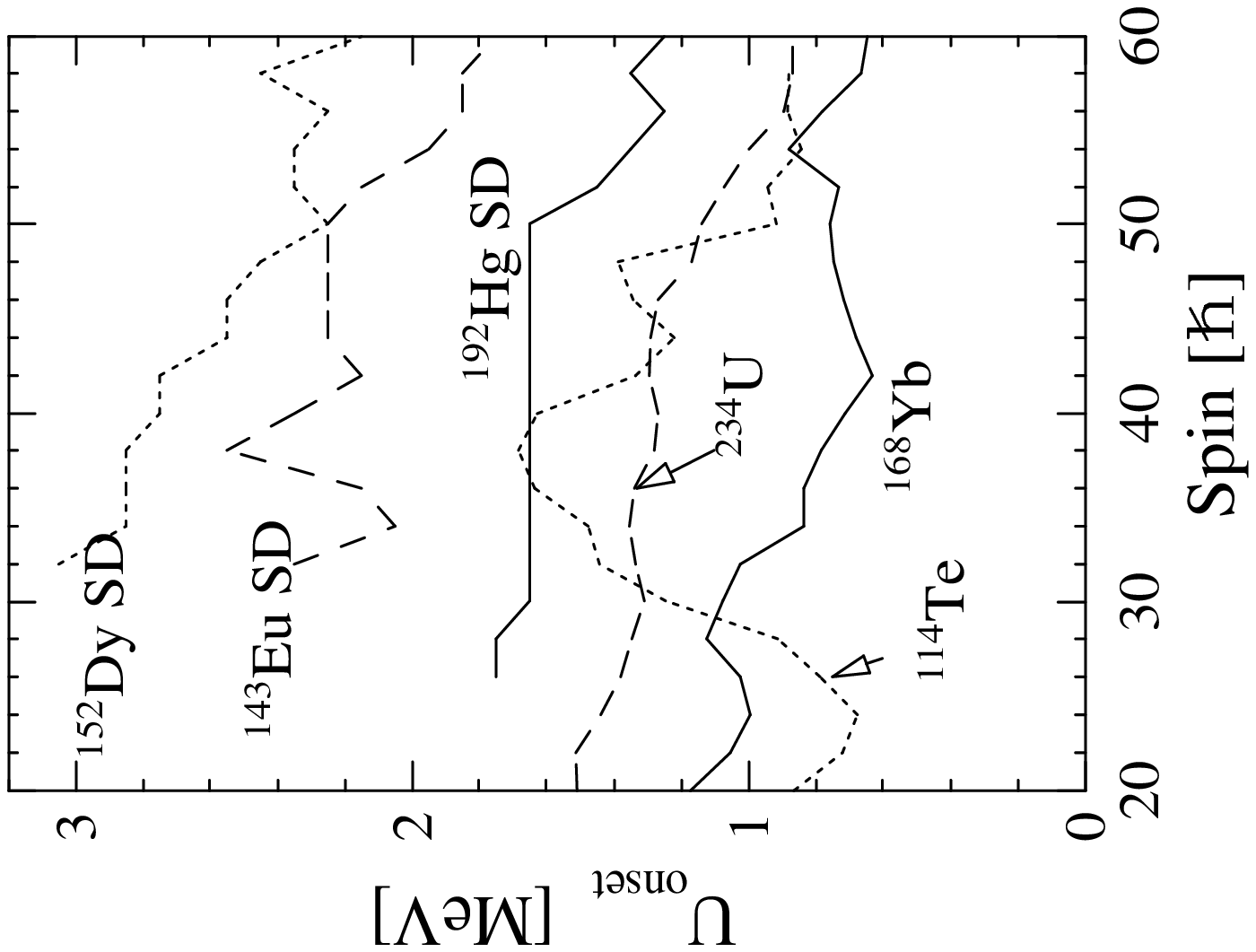,height=7.5cm,angle=-90}}
\caption{The number of rotational bands \Nband\ and the
onset energy $U_{onset}$ of rotational damping for
normal deformed${}^{114}$Te, ${}^{168}$Yb, \U\ and superdeformed
${}^{143}$Eu,${}^{152}$Dy, ${}^{192}$Hg, as a function of spin. 
}
\end{figure}

The extracted spreading width \Gammamu\ of np-nh cranked shell model
states is also listed in Table 1 together
with \Gammarot. Compared with the large variation in \Gammarot, 
the spreading width takes more or less similar value 
for different mass regions. It is noted that  \Gammarot\ 
is larger than the spreading width \Gammamu\ in \Yb, which 
implies the weak
coupling situation for which the concept of doorway states is 
meaningful. Accordingly the gamma-gamma correlation spectrum
displays the two-component
profile. The weak coupling situation holds also for \Te\ and SD
\Eu. On the contrary, the spreading width \Gammamu\ is comparable with or
slightly smaller than \Gammarot\ for superdeformed 
\Hg, \Dy, and normal deformed \U. If $\Gamma_\mu \ll \Gamma_{rot}$ 
held, the ``motional narrowing'' 
would be expected. SD \Hg, SD \Dy, and \U\ are situated at the borderline for
onset of the motional narrowing.

A variety of features is also seen considering the 
onset of the rotational damping.
Figure 6 shows the excitation energy \Uonset (measured
from the yrast line) where the damping sets in, and the
number \Nband\ of rotational bands  which lie near the yrast line,
surviving against the rotational damping
\cite{Matsuo97,Yoshida97,Yoshida98}.
Compared to the typical example \Yb\ of rare-earth normal deformed
nuclei, the onset energy is  higher in
superdeformed \Eu\  and \Dy\ by more than 1 MeV. 
 This is because the level density of superdeformed states
is much lower than in normal deformed nuclei due to the
shell gap in the cranked single-particle spectrum\cite{Aberg,Yoshida97}. 
Also the value of \Nband\ is larger than in
$^{168}$Yb.
It is noted
that \Nband\ for  superdeformed \Hg\ reaches 150. In SD \Hg\ 
the level density is not as small as in other SD nuclei, and the
anomalously large value of \Nband\ is not explained by the level density
effect. Instead, this is caused by 
the extremely small rotational damping width
 \Gammarot$ \sim 20$ keV, which is comparable
with the level spacing ($d=30-10$ keV) in the onset region
 ($E=1.2-1.6$ MeV)\cite{Yoshida98}. 
If $\Gamma_{rot} \ll d$, the configuration mixing
among np-nh cranked shell model states should not lead to
the damping of rotation, as was discussed by Mottelson \cite{Mottelson};
one should instead expect rotational band structures built upon 
the strongly mixed
compound states, the so called ``ergodic rotational
bands'' \cite{Aberg96}. 
The case of SD \Hg\ is close to the borderline
for the emergence of such ergodic rotational bands \cite{Yoshida98}.

\end{document}